\documentstyle[aps,multicol,prb]{revtex}

\begin{document}

\draft

\title{Subtleties in Quantum Mechanical Metastability}
\author{R. M. Cavalcanti\cite{rmc}}
\address{Departamento de F\'\i sica,
Pontif\'\i cia Universidade Cat\'olica do Rio de Janeiro, \\
Cx. Postal 38071, CEP 22452-970, Rio de Janeiro, RJ, Brazil}
\author{C. A. A. de Carvalho\cite{aragao}}
\address{Instituto de F\'\i sica,
Universidade Federal do Rio de Janeiro, \\
Cx. Postal 68528, CEP 21945-970, Rio de Janeiro, RJ, Brazil}
\date{April 28, 1997}
\maketitle

\begin{abstract}

We present a detailed discussion of some features of quantum mechanical
metastability. We analyze the nature of decaying (quasistationary) states
and the regime of validity of the exponencial law, as well as decays at
finite temperature. We resort to very simple systems and elementary 
techniques to emphasize subtle aspects of the problem.

\end{abstract}

\multicols{2}

\section{Introduction}
\label{I}

Complex-energy eigenfunctions made their d\'ebut in Quantum
Mechanics through the hands of Gamow, in the theory of alpha-decay.
\cite{Gamow} Gamow imposed an ``outgoing wave boundary condition''
on the solutions of the Schr\"odinger
equation for an alpha-particle trapped in the nucleus. Since 
there is only an {\em outgoing} flux of alpha-particles, the wave
function $\psi(r,t)$ must behave far from the nucleus as 
(for simplicity, we consider an s-wave)\cite{units}
\begin{equation}
\label{1}
\psi(r,t)\sim\frac{e^{-iEt+ikr}}{r}\qquad(r\rightarrow\infty).
\end{equation}
This boundary condition, together with the requirement
of finiteness of the wave function at the origin, gives rise
to a quantization condition on the values of $k$ (and,
therefore, on the values of $E=k^2/2$). It turns out
that such values are complex:
\begin{equation}
\label{2}
k_n=\kappa_n-iK_n/2,\qquad E_n=\epsilon_n-i\Gamma_n/2.
\end{equation}
It follows that
\begin{equation}
\label{3}
|\psi_n(r,t)|^2\sim\frac{e^{-\Gamma_nt+K_nr}}{r^2}
\qquad(r\rightarrow\infty).
\end{equation}
Thus, if $\Gamma_n>0$, the probability of finding the alpha-particle
in the nucleus decays exponentially in time. The lifetime of the 
nucleus would be given by $\tau_n=1/\Gamma_n$, and the energy
of the emitted alpha-particle by $\epsilon_n$.

Although very natural, this interpretation suffers from
some difficulties. How can the energy, which is an observable quantity,
be complex? In other words, how can the Hamiltonian, which is a
Hermitean operator, have complex eigenvalues? Also, the eigenfunctions
are not normalizable, since $\Gamma_n$ positive implies $K_n$ positive and,
therefore, according to (\ref{3}),
$|\psi_n(r,t)|^2$ diverges exponentially with $r$.

In spite of such problems (which, in fact, are closely related),
it is a fact of life that alpha-decay, as well as other
types of decay, does obey an exponential decay law, with a rate 
close to that obtained using Gamow's method. Why this method works
is the question we try to answer in this paper in a very elementary
way.\cite{Holstein} Thus, in Section \ref{II},
we show Gamow's method in action for a very simple potential. Some
of the results obtained there are used in Section \ref{III}, where we
study the time evolution of a wave packet initially confined
in the potential well defined in Section \ref{II}. This is done
with the help of the propagator built with normalizable\cite{delta}
eigenfunctions, associated to real eigenenergies.
As a bonus, we show that the
exponential decay law is not valid for very small times or
for very large times. This is the content of Section \ref{IV}, where
the region of validity of the exponential decay law is roughly delimited.

Another topic we address in this paper is decay at finite
temperature. This is done in Section \ref{V}, where we study, with the 
help of an exactly solvable toy model, how the decay of a 
metastable system is modified when it is coupled to a heat bath. 
It is shown that, under suitable conditions, the decay is exponential,
with a decay rate $\Gamma$ given by the thermal average
of the $\Gamma_n$'s,
\begin{equation}
\label{3.5}
\Gamma=\frac{\sum_n\Gamma_n\,e^{-\beta \epsilon_n}}
{\sum_n e^{-\beta \epsilon_n}}.
\end{equation}
Although this result appears to be rather obvious, in fact it is
not: the decay of a metastable system is an intrinsically
non-equilibrium process and, so, there is no a priori reason for
the decay rate to be given by (\ref{3.5}). Finally, in Section \ref{VI}, we 
discuss the concept of ``free energy of a metastable phase,'' a
point where we think there is some confusion in the literature.

The results and ideas presented here are not really new, but 
discussions on these matters usually involve the use of
sophisticated mathematical techniques, such as functional\cite{Skib,BGB} 
or complex analysis.\cite{Beck & Nuss,Nuss,CSM,FGR,Moshinsky}
For this reason, we have tried to make the presentation as clear
as possible
by resorting to very simple systems and elementary techniques ---
in fact, techniques that can be found in any standard Quantum
Mechanics textbook.\cite{Merzbacher}

\section{Decaying States}
\label{II}

In order to exhibit Gamow's method in action, we shall study
the escape of a particle from the potential well given by:
\begin{equation}
\label{4}
V(x)=\cases{(\lambda/a)\,\delta(x-a) & for $x>0$, \cr
+\infty & for $x<0$. \cr}
\end{equation}
Motion in the region $x<0$ is forbidden because of the infinite wall
at the origin. The positive dimensionless constant $\lambda$
is a measure of the ``opacity'' of the barrier at $x=a$; in the limit
$\lambda\rightarrow\infty$, the barrier becomes impenetrable,
and the energy levels inside the well are quantized. If $\lambda$
is finite, but very large, a particle is no more confined to the
well, but it usually stays there for a long time before it escapes.
 If $\lambda$ is not so large, the particle can easily
tunnel through the barrier, and quickly escape from
the potential well. Metastability, therefore, can only be achieved
if the barrier is very opaque, i.e., $\lambda$ is very large.
For this reason, we shall assume this to be the case in what follows
and, whenever possible, we shall retain only the first non-trivial
term in a $1/\lambda$ expansion.

To find out how fast the particle escapes from the potential
well, we must solve the Schr\"odinger equation 
\begin{equation}
\label{5}
i\,\frac{\partial}{\partial t}\,\psi(x,t)=-\frac{1}{2}\,
\frac{\partial^2}{\partial x^2}\,\psi(x,t)+\frac{\lambda}{a}\,
\delta(x-a)\,\psi(x,t).
\end{equation}
$\psi(x,t)=\exp(-iEt)\,\varphi(x)$ is a particular solution
of this equation, provided $\varphi(x)$ satisfies the time-independent
Schr\"odinger equation
\begin{equation}
\label{6}
-\frac{1}{2}\,\frac{d^2}{d x^2}\,\varphi(x)+\frac{\lambda}{a}\,
\delta(x-a)\,\varphi(x)=E\,\varphi(x).
\end{equation}
Denoting the regions $0<x<a$ and $x>a$ by the indices 1 and 2,
respectively, the corresponding wave functions $\varphi_j(x)$ $(j=1,2)$
satisfy the free-particle Schr\"odinger equation:
\begin{equation}
\label{7}
-\frac{1}{2}\,\frac{d^2}{d x^2}\,\varphi_j(x)=E\,\varphi_j(x).
\end{equation}
Since the wall at the origin is impenetrable, $\varphi_1(0)$ must
be zero; the solution of Eq.\ (\ref{7}) which obeys this boundary
condition is
\begin{equation}
\label{8}
\varphi_1(x)=A\,\sin kx\qquad(k=\sqrt{2E}\,).
\end{equation}
To determine $\varphi_2(x)$, we follow Gamow's reasoning\cite{Gamow,BGB,Gold}
and require $\varphi_2(x)$ to be an outgoing wave.
Therefore, we select, from the admissible solutions of Eq.\ (\ref{7}),
\begin{equation}
\label{9}
\varphi_2(x)=B\,e^{ikx}.
\end{equation}
The wave function must be continuous at $x=a$, so that $\varphi_1(a)=
\varphi_2(a)$, or
\begin{equation}
\label{10}
\frac{B}{A}=e^{-ika}\,\sin ka.
\end{equation}
On the other hand, the derivative of the wave function has a
discontinuity at $x=a$, which can be determined by integrating both
sides of (\ref{6}) from $a-\varepsilon$ to $a+\varepsilon$, with
$\varepsilon\rightarrow 0^+$:
\begin{equation}
\label{11}
\varphi_2'(a)-\varphi_1'(a)=\frac{2\lambda}{a}\,\varphi_2(a),
\end{equation}
from which there follows another relation between $A$ and $B$:
\begin{equation}
\label{12}
\frac{B}{A}=\frac{e^{-ika}\,\cos ka}{i-2\lambda/ka}.
\end{equation}
Combining (\ref{10}) and (\ref{12}), we obtain a quantization condition
for $k$:
\begin{equation}
\label{13}
{\rm cotan}\,ka=i-\frac{2\lambda}{ka}.
\end{equation}
The roots of Eq.\ (\ref{13}) are complex; 
when $\lambda\gg 1$, those which are closest to the origin
are given by\cite{Nuss}
\begin{eqnarray}
\label{14}
k_n&\approx& \frac{1}{a}\left[\frac{2n\pi\lambda}{1+2\lambda}-
i\left(\frac{n\pi}{2\lambda}\right)^2\right]
\qquad(n\pi\ll\lambda) \\
&\equiv& \kappa_n-iK_n/2. \nonumber
\end{eqnarray}
The corresponding eigenenergies are
\begin{eqnarray}
\label{15}
E_n&=&k_n^2/2\approx\frac{1}{2}\left(\frac{n\pi}{a}
\right)^2-i\,\frac{(n\pi)^3}{(2\lambda a)^2} \\
&\equiv&\epsilon_n-i\Gamma_n/2. \nonumber
\end{eqnarray}
(Note that $K_n\ll\kappa_n$ and $\Gamma_n\ll\epsilon_n$; these
results will be used later.)
The imaginary part of $E_n$ gives rise to an exponential decay
of $|\psi_n(x,t)|^2$, with lifetime equal to
\begin{equation}
\label{16}
\tau_n=1/\Gamma_n\approx\frac{2(\lambda a)^2}{(n\pi)^3}.
\end{equation}
Since the corresponding value of $B/A$ is very small ($\sim n/\lambda$),
one may be tempted to say that the probability of finding the particle
outside the well is negligible in comparison with the probability
of finding the particle inside the well. Normalizing $\psi_n$ in such 
a way that the latter equals one when $t=0$, the probability
of finding the particle inside the well at time $t$, if it were in the
$n$-th decaying state at $t=0$, would be
\begin{equation}
\label{17}
P_n(t)=\int_0^a|\psi_n(x,t)|^2\,dx=\exp(-\Gamma_nt).
\end{equation}

The trouble with this interpretation is that ${\rm Im}\,k_n\equiv-K_n/2<0$,
and so $\psi_n(x,t)$ diverges exponentially as $x\rightarrow\infty$, since,
according to (\ref{9}), 
\begin{equation}
\label{18}
|\psi_n(x,t)|^2=|B_n|^2\,\exp(-\Gamma_nt+K_nx)
\end{equation}
outside the well. Because of this ``exponential catastrophe'', the
decaying states are nonnormalizible and, therefore, cannot be
accepted as legitimate solutions of the 
Schr\"odinger equation (although one can find in the literature the
assertion that they are ``rigorous'' solutions of the time-dependent
Schr\"odinger equation\cite{Legget}). 

\section{Time Evolution of a Wave Packet}
\label{III}

We now return to Eq.\ (\ref{7}) and write, for the solution in region 2,
instead of (\ref{9}), the sum of an outgoing plus an incoming wave:
\begin{equation}
\label{19}
\varphi_2(x)=e^{-ikx}+B\,e^{ikx}.
\end{equation}
Continuity of the wave function at $x=a$ implies
\begin{equation}
\label{20}
A\,\sin ka=e^{-ika}+B\,e^{ika}.
\end{equation}
As before, the derivative of the wave function has a discontinuity
at $x=a$, given by Eq.\ (\ref{11}), from which it follows, instead of
(\ref{12}),
\begin{equation}
\label{21}
kA\,\cos ka=-\left(\frac{2\lambda}{a}+ik\right)e^{-ika}
-\left(\frac{2\lambda}{a}-ik\right)B\,e^{ika}.
\end{equation}
Solving (\ref{20}) and (\ref{21}) for $A$ and $B$, we find
\begin{mathletters}
\label{22}
\begin{eqnarray}
A&=&-\frac{2ika}{(ka+\lambda\,\sin 2ka)+i\,2\lambda\,\sin^2 ka}, \\
B&=&-\frac{(ka+\lambda\,\sin 2ka)-i\,2\lambda\,\sin^2 ka}
{(ka+\lambda\,\sin 2ka)+i\,2\lambda\,\sin^2 ka}.
\end{eqnarray}
\end{mathletters}

These expressions show a couple of interesting features: 

(1) $|B|=1$ for
real values of $k$, implying a zero net flux of probability through
$x=a$; therefore, unlike the solution found
in the previous section, there is no loss or accumulation of 
probability in the well region. 

(2) $|A|\ll 1$ if $ka\ll\lambda$, except if $k$ is close to
a pole of $A(k)$, in which case $|A|$ may become very large.

To find the poles of $A$ we must solve the equation $A(k)^{-1}=0$, 
which, after some algebraic manipulations, reads
\begin{equation}
\label{23}
{\rm cotan}\,ka=i-\frac{2\lambda}{ka}.
\end{equation}
This is the same as Eq.\ (\ref{13})! Is this a coincidence? In fact, no.
According to (\ref{22}), $A$ and $B$ have the same 
poles; in a sufficiently small vicinity of a pole, $|A|$ and $|B|$ are
very large, and so Eqs.\ (\ref{20}) and (\ref{21}) become
equivalent to Eqs.\ (\ref{10}) and (\ref{12}), respectively.

Suppose that at $t=0$ the particle is known to be in
the region $x<a$ with probability 1; in other words, its wave function
$\psi(x,0)$ is zero outside the well,
\begin{equation}
\label{24}
\psi(x,0)=0\qquad{\rm for}\;x>a.
\end{equation}
The wave function at a later time $t$ is given by
\begin{equation}
\label{25}
\psi(x,t)=\int_0^{\infty}G(x,x';t)\,\psi(x',0)\,dx',
\end{equation}
where the propagator, $G(x,x';t)$, can be written as
\begin{equation}
\label{26}
G(x,x';t)=\int_0^{\infty}e^{-ik^2t/2}\,\varphi^*(k,x)
\,\varphi(k,x')\,dk.
\end{equation}
The function $\varphi(k,x)$ is the solution of Eq.\ (\ref{6})
corresponding to the energy $E=k^2/2$:
\begin{equation}
\label{27}
\varphi(k,x)=\frac{1}{\sqrt{2\pi}}\times\cases{
A(k)\,\sin kx & $(x<a)$ \cr
e^{-ikx}+B(k)\,e^{ikx} & $(x>a)$. \cr}
\end{equation}
With this normalization, the $\varphi(k,x)$ satisfy the completeness
relation\cite{Landau}
\begin{equation}
\label{28}
\int_0^{\infty}\varphi^*(k,x)\,\varphi(k,x')\,dk=\delta(x-x').
\end{equation}
Since, by hypothesis, $\psi(x,0)=0$ for $x>a$, (\ref{25})--(\ref{27})
give, for $x<a$,
\begin{eqnarray}
\label{29}
\psi(x,t)&=&\frac{1}{2\pi}\int_0^{\infty}dk\,e^{-ik^2t/2}\,
|A(k)|^2\,\sin kx\, \nonumber \\
&\times&\int_0^a dx'\,\psi(x',0)\,\sin kx'.
\end{eqnarray}
For $k$ close to a pole $k_n\equiv \kappa_n-iK_n/2$, $A(k)$ can be approximated by
\begin{eqnarray}
\label{30}
A(k)&\approx& -\frac{2ik_na}{a\,(1+2\lambda\,e^{2ik_na})\,
(k-k_n)} \nonumber \\
&\approx& -\frac{i\kappa_n/\lambda}{(k-\kappa_n)+iK_n/2}.
\end{eqnarray}
As we have seen, $|A(k)|^2\ll 1$ if $ka\ll\lambda$, except at the 
resonances, where
(\ref{30}) may be used. On the other hand, if $\psi(x,0)$ is 
sufficiently smooth, in the sense that
$\int_0^a dx\,\psi(x,0)\,\sin kx\rightarrow 0$
sufficiently fast when $k\rightarrow\infty$ (this condition will be
made more precise later),
then most of the contribution to the integral (\ref{29}) comes
from the region $ka\ll\lambda$.
Therefore, we may approximate (\ref{29}) by
\begin{eqnarray}
\label{31}
\psi(x,t)&\approx& \frac{1}{2\pi}\sum_{n}\int_{I_n}
dk\,e^{-ik^2t/2}\,\frac{(\kappa_n/\lambda)^2\,
\sin kx}{(k-\kappa_n)^2+K_n^2/4}\, \nonumber \\
&\times&\int_0^a dx'\,\psi(x',0)\,\sin kx',
\end{eqnarray}
where $I_n$ is the interval $[(\kappa_n+\kappa_{n-1})/2,
(\kappa_n+\kappa_{n+1})/2]$ $(n=1,2,\ldots; \kappa_0\equiv 0)$.
Because of the arguments preceding (\ref{31}), only the first few
terms of the sum give a significant contribution to the integral.
Note also that, since the resonance in $|A(k)|^2$ around $\kappa_n$ has
a width of the order of $K_n$, and
\begin{equation}
\label{32}
K_nx,\,K_nx'\le K_na\approx (n\pi)^2/2\lambda^2\ll 1,
\end{equation}
we can substitute $k$ for $\kappa_n$ in $\sin kx$ and $\sin kx'$ in
the integrand of (\ref{31}). On the other hand, this is not
allowed for $e^{-ik^2t/2}$, since the time $t$ is not bounded.

Let us examine the integrals
\begin{equation}
\label{33}
{\cal I}_n(t)\equiv\int_{I_n}dk\,e^{-ik^2t/2}\,
\frac{1}{(k-\kappa_n)^2+K_n^2/4}.
\end{equation}
If $K_n\ll \kappa_{n+1}-\kappa_{n-1}$, we can safely extend the interval of
integration to the whole real axis, and carrying out the integration
we find
\begin{mathletters}
\label{34}
\begin{eqnarray}
\label{34a}
{\cal I}_n(t)&\approx&\int_{-\infty}^{\infty}dk\,e^{-ik^2t/2}\,
\frac{1}{(k-\kappa_n)^2+K_n^2/4} \nonumber \\
&=&\frac{1}{\sqrt{2\pi it}}\,\int_{-\infty}^{\infty}d\xi\,
e^{i\xi^2/2t}\,\int_{-\infty}^{\infty}\frac{dk\,e^{i\xi k}}
{(k-\kappa_n)^2+K_n^2/4} \nonumber \\
&=&\frac{2\pi}{K_n\,\sqrt{2\pi it}}\,({\cal I}_n^1+{\cal I}_n^2); \\
{\cal I}_n^1&\equiv&\int_{-\infty}^{0}d\xi\,e^{i\xi^2/2t+
i(\kappa_n-iK_n/2)\xi}, \\
{\cal I}_n^2&\equiv&\int_{0}^{\infty}d\xi\,e^{i\xi^2/2t+
i(\kappa_n+iK_n/2)\xi}. 
\end{eqnarray}
\end{mathletters}

\noindent
Except for a region of width $\Delta\xi\sim\sqrt{t}$ around the point
$\xi=-\kappa_nt$, where the phases of the exponentials are stationary,
the oscillations of the integrands tend to cancel out, giving a
very small contribution to the integrals above. If $\kappa_nt\gg\sqrt{t}$,
such a region is well inside the negative real axis, therefore the
second integral can be neglected in comparison to the first. For the
same reason, we can extend the interval of integration of the first
integral to the whole real axis, thus obtaining
\begin{eqnarray}
\label{35}
{\cal I}_n(t)&\approx&\frac{2\pi}{K_n\,\sqrt{2\pi it}}\,
\int_{-\infty}^{\infty}d\xi\,e^{i\xi^2/2t+i(\kappa_n-iK_n/2)\xi}
\nonumber \\
&=&\frac{2\pi}{K_n}\,e^{-i(\kappa_n-iK_n/2)^2t/2}
=\frac{2\pi}{K_n}\,e^{-iE_nt}.
\end{eqnarray}
Substituting this result in Eq.\ (\ref{31}), we find
\begin{equation}
\label{36}
\psi(x,t)\approx\sum_{n}c_n\,e^{-iE_nt}\,\varphi_n(x),
\end{equation}
where $\varphi_n(x)$ and $c_n$ are defined as
\begin{mathletters}
\label{37}
\begin{eqnarray}
\varphi_n(x)&=&\sqrt{\frac{a}{2}}\,\frac{(\kappa_n/\lambda)^2}{K_n}\,
\sin \kappa_n x
\approx\sqrt{\frac{2}{a}}\,\sin \frac{n\pi x}{a}, \\
c_n&=&\int_0^a dx\,\psi(x,0)\,\varphi_n(x).
\end{eqnarray}
\end{mathletters}

\noindent
Eq.\ (\ref{36}) is formally identical to the well known
expansion of the wave function in energy eigenfunctions, except
for the fact that: (1) it is an {\em approximate} result and, as
such, subject to some restrictions, and (2) the energies $E_n$
are complex, as a result of which the probability $P(t)$ to find the
particle inside the potential well at time $t$ decreases in time:
\begin{equation}
\label{38}
P(t)=\int_0^a|\psi(x,t)|^2\,dx\approx\sum_{n}|c_n|^2\,
e^{-\Gamma_n t}.
\end{equation}
Now we can be more precise on the smoothness of $\psi(x,0)$; roughly
speaking, the smaller the value of $n$ beyond which $c_n=0$, the
better the results above will be.

Eq.\ (\ref{36}) is valid only inside the potential well. 
To find the wave function outside the well, we must return to
Eqs.\ (\ref{25})--(\ref{27}) and make $x>a$:
\begin{eqnarray}
\label{39}
\psi(x,t)&=&\frac{1}{2\pi}\,\int_0^{\infty}dk\,e^{-ik^2t/2}\,
[e^{ikx}+B^*(k)\,e^{-ikx}]\,A(k)\, \nonumber \\
&\times&\int_0^a dx'\,\psi(x',0)\,\sin kx'.
\end{eqnarray}
Since most of the contribution to the integral in $k$ comes from
the resonances, we may approximate $A(k)$ by (\ref{30}) and make
an analogous approximation for $B^*(k)$. Thus, (\ref{39}) becomes
\begin{mathletters}
\label{40}
\begin{eqnarray}
\psi(x,t)&\approx&\sum_{n}d_n\,{\cal J}_n(x,t); \\
d_n&\equiv&\frac{-i\kappa_n}{2\pi\lambda}\,\int_0^a dx'\,\psi(x',0)\,
\sin \kappa_nx', \\
{\cal J}_n(x,t)&\equiv&\int_{I_n}dk\,
e^{-ik^2t/2} \left(\frac{e^{ikx}}{k-\kappa_n+iK_n/2}-{\rm c.c.}\right).
\nonumber \\
\end{eqnarray}
where $I_n$ has the same meaning as in Eq.\ (\ref{31}), and c.c.\ denotes
complex conjugate.
\end{mathletters}

Let us concentrate our attention on the integrals ${\cal J}_n(x,t)$.
Extending the interval of integration to the whole real axis,
and using the same trick as in Eq.\ (\ref{34a}), we find
\begin{mathletters}
\label{41}
\begin{eqnarray}
{\cal J}_n(x,t)&\approx&-\frac{2\pi i}{\sqrt{2\pi it}}\,
({\cal J}_n^1+{\cal J}_n^2); \\
{\cal J}_n^1&\equiv&\int_{-\infty}^{-x}d\xi\,e^{i\xi^2/2t+ik_n(\xi +x)}, \\
{\cal J}_n^2&\equiv&\int_{x}^{\infty}d\xi\,e^{i\xi^2/2t+ik_n^*(\xi -x)}.
\end{eqnarray} 
\end{mathletters}

\noindent
As in the case of ${\cal I}_n(t)$, the second integral is negligible
in comparison to the first if $\kappa_nt\gg\sqrt{t}$, or $t\gg 1/\epsilon_n$.
On the other hand, the first integral may be approximated by
\begin{equation}
\label{42}
\int_{-\infty}^{\infty}d\xi\,e^{i\xi^2/2t+ik_n(\xi +x)}
=\sqrt{2\pi it}\,e^{-ik_n^2t/2+ik_nx}
\end{equation}
only if $\kappa_nt-x\ll\sqrt{t}$. 

Returning to Eq.\ (\ref{40}), we finally obtain
\begin{equation}
\label{43}
\psi(x,t)\approx -2\pi i\sum_{n}d_n\,e^{-iE_nt+ik_nx}.
\end{equation} 
We see, therefore, that outside the well the wavefunction behaves as
a superposition of outgoing waves, in the way postulated by Gamow.
However, the exponential catastrophe does not occur here, for
Eq.\ (\ref{43}) is valid only under the assumption that
$\kappa_nt-x\ll\sqrt{t}$.

\section{Breakdown of Exponential Decay}
\label{IV}

The evolution of the wave function requires some time\cite{initial}
to reach the
regime of exponential decay; typically, a time corresponding to
many oscillations inside the potential well (i.e., $t\gg 1/\epsilon_n$).
To be more precise, even if this condition is satisfied, the decay
is not strictly exponential, but a sum of exponential decays, one
for each resonance [Eq.\ (\ref{38})]. However, since the lifetime
$\tau_n$ is, in general, a rapidly decreasing function of $n$
($\tau_n\approx\tau_1/n^3$ in our example), the decay becomes a pure
exponential one after a time of the order of $\tau_1$. 

On the other hand, the exponential decay does not last forever.
After some sufficiently long time, it obeys a power 
law.\cite{Beck & Nuss,Nuss,CSM,FGR,Moshinsky,Patra}
To see this, note that for $t\rightarrow\infty$, the integral
(\ref{29}) is dominated by small values of $k$. One finds, then,
for $x<a$,
\begin{eqnarray}
\label{44}
\psi(x,t)&\approx&\frac{|A(0)|^2\,x}{2\pi}\,\int_0^a dx'\,
\psi(x',0)\,x'\,\int_0^{\infty}dk\,e^{-ik^2t/2}\,k^2
\nonumber \\
&=&\frac{x}{\lambda^2\,\sqrt{8\pi it^3}}\,\int_0^adx'\,\psi(x',0)\,
x'. 
\end{eqnarray}
Therefore, the probability of finding the particle inside the
potential well behaves asymptotically as
\begin{equation}
\label{45}
P(t)\approx\frac{a^3}{24\pi\lambda^4t^3}\,
\left|\,\int_0^a dx'\,\psi(x',0)\,x'\,\right|^2
\sim\frac{a^6}{\lambda^4 t^3}.
\end{equation}
Comparing (\ref{45}) with (\ref{38}), one finds
that they become comparable in magnitude when
\begin{equation}
\label{46}
e^{-t/\tau_1}\sim\frac{a^6}{\lambda^4 t^3}\sim\lambda^{-10}\,
\left(\frac{\tau_1}{t}\right)^{-3},
\end{equation}
or, since $\lambda\gg 1$,
\begin{equation}
\label{47}
\frac{t}{\tau_1}\sim 10\,\ln \lambda +3\,\ln \ln \lambda.
\end{equation}
Thus, when the decay begins to obey a power law\cite{Ricardo} 
$(\sim t^{-3})$,
the probability that the particle is still inside the potential
well is so small $(\lesssim\lambda^{-10})$, that it would be very
difficult to observe deviations from the exponential decay.

\section{Decay at Finite Temperature}
\label{V}

In general, the initial state of the particle, $\psi(x,0)$, is not
precisely known. Such a knowledge is required in order to determine
the coeficients $c_n$ in Eq.\ (\ref{36}). Let us imagine, however,
that the system is in contact with a heat bath at temperature $T$.
Then, it is reasonable to assume that ($\beta=1/k_BT$, $k_B=$ Boltzmann
constant)
\begin{equation}
\label{48}
|c_n|^2=\frac{e^{-\beta \epsilon_n}}{Z},\qquad
Z\equiv\sum_n e^{-\beta \epsilon_n}.
\end{equation}
Thus, Eq.\ (\ref{36}) gives
\begin{equation}
\label{49}
P(t)\approx\frac{1}{Z}\,\sum_n e^{-\beta \epsilon_n-\Gamma_n t},
\end{equation}
and, as already discussed in Sec.\ \ref{IV}, after a time of the order
$\tau_1=1/\Gamma_1$, the decay would be dominated by the decay
of the ``false vacuum'' --- the lowest lying resonance.
It follows that the decay rate is almost insensitive to the
temperature.

Is this conclusion correct?

In the literature\cite{Affleck} one finds the statement that,
in such a situation, the probability $P(t)$ decays as
\begin{equation}
\label{50}
P(t)=e^{-\langle\Gamma\rangle t},\qquad
\langle\Gamma\rangle=\frac{1}{Z}\,\sum_n\Gamma_n\,e^{-\beta \epsilon_n}.
\end{equation}
A perfectly sensible question is: why does $P(t)$ decay this way,
and not as
\begin{equation}
\label{52}
P(t)=e^{-t/\langle\tau\rangle},\qquad
\langle\tau\rangle=\frac{1}{Z}\,\sum_n\tau_n\,e^{-\beta \epsilon_n}\,?
\end{equation}

In order to answer these questions, we shall study a toy model:
a two-level metastable system coupled to a heat bath. (For instance,
imagine a situation in which only the two lowest resonances of a
potential well are excited.) These levels have
``complex energies'' $E_j=\epsilon_j-i\Gamma_j/2$ ($j=1,2$). 
Let $n_1(t)$ and $n_2(t)$ be the populations at
time $t$ of levels 1 and 2, respectively. A reasonable dynamics is given
by the following set of equations ($\dot{n}\equiv dn/dt$, 
$E\equiv \epsilon_2-\epsilon_1$):
\begin{equation}
\label{53}
\cases{\dot{n}_1=-\Gamma_1\,n_1+\Gamma\,(n_2-e^{-\beta E}\,n_1), \cr
\dot{n}_2=-\Gamma_2\,n_2-\Gamma\,(n_2-e^{-\beta E}\,n_1). \cr}
\end{equation}
The first term on the r.h.s.\ of (\ref{53}) describes the ``natural''
decay of the levels. The second term is due to the coupling to the
the heat bath; it drives the system towards
thermal equilibrium with the bath.

Since (\ref{53}) is a set of differential linear equations,
solutions can be found in the form
\begin{equation}
\label{54}
n_1(t)=a\,e^{-\lambda t},\qquad n_2(t)=b\,e^{-\lambda t}.
\end{equation}
The decay rate, $\lambda$, must satisfy the characteristic equation
\begin{eqnarray}
\label{55}
\lambda^2-[\Gamma_1+\Gamma_2&+&(1+e^{-\beta E})\Gamma]\,\lambda \nonumber \\
&+&\Gamma_1\Gamma_2+(\Gamma_1+e^{-\beta E}\,\Gamma_2)\Gamma=0.
\end{eqnarray}
Although this equation can be solved exactly, the exact solution
is not very illuminating. Instead, we shall consider two limiting
situations.

(1) $\Gamma\gg\Gamma_1$, $\Gamma_2$ (overdamping): in this case,
we may approximate Eq.\ (\ref{55}) by
\begin{equation}
\label{56}
\lambda^2-(1+e^{-\beta E})\Gamma\lambda
+(\Gamma_1+e^{-\beta E}\,\Gamma_2)\Gamma=0,
\end{equation}
whose solutions are
\begin{equation}
\label{57}
\lambda_{\pm}=\frac{1}{2}\,(1+e^{-\beta E})\Gamma
\left(1\pm\sqrt{1-\frac{4(\Gamma_1+e^{-\beta E}\,\Gamma_2)}
{(1+e^{-\beta E})^2\,\Gamma}}\,\right).
\end{equation}
Expanding the square root, we find (since $\Gamma_1$, $\Gamma_2\ll
\Gamma$)
\begin{mathletters}
\label{58}
\begin{eqnarray}
\lambda_+&\approx&(1+e^{-\beta E})\Gamma, \\
\lambda_-&\approx&\frac{\Gamma_1+e^{-\beta E}\,\Gamma_2}
{1+e^{-\beta E}}\equiv\langle\Gamma\rangle.
\end{eqnarray}
\end{mathletters}

\noindent
The corresponding eigenvectors are
\begin{equation}
\label{59}
\pmatrix{a_+\cr b_+\cr}\approx\pmatrix{1\cr -1\cr},\qquad
\pmatrix{a_-\cr b_-\cr}\approx\pmatrix{1\cr e^{-\beta E}\cr}.
\end{equation}
If $n_j(0)=n_{j0}$ $(j=1,2)$, then, at time $t$, we have
\begin{eqnarray}
\label{60}
\pmatrix{n_1(t)\cr n_2(t)\cr}&\approx&\frac{e^{-\beta E}\,n_{10}-n_{20}}
{1+e^{-\beta E}}\,\pmatrix{1\cr -1\cr}e^{-\lambda_+t} \nonumber \\
&+&\frac{n_{10}+n_{20}}{1+e^{-\beta E}}\,\pmatrix{1\cr e^{-\beta E}\cr}
e^{-\lambda_-t}.
\end{eqnarray}
Therefore, even if the system is initially not in thermal equilibrium
with the heat bath, it thermalizes in a time of the order $1/\lambda_+
\ll 1/\lambda_-$. In other words, after a transient time of the order
of $1/\lambda_+$, both levels decay at the same
rate, $\lambda_-$, equal to the thermal average of $\Gamma_1$ and $\Gamma_2$,
and their populations soon approach the Boltzmann distribution,
\begin{equation}
\label{61}
\frac{n_2(t)}{n_1(t)}\approx e^{-\beta E}.
\end{equation}

(2) $\Gamma\ll\Gamma_1$, $\Gamma_2$ (underdamping): now, Eq.\ (\ref{55})
may be approximated by
\begin{equation}
\label{62}
\lambda^2-(\Gamma_1+\Gamma_2)\lambda+\Gamma_1\Gamma_2=0,
\end{equation}
whose solutions are $\lambda_j=\Gamma_j$ $(j=1,2)$.
The corresponding eigenvectors are\cite{eigenvector}
\begin{equation}
\label{63}
\pmatrix{a_1\cr b_1\cr}\approx\pmatrix{1\cr 0\cr},\qquad 
\pmatrix{a_2\cr b_2\cr}\approx\pmatrix{0\cr 1\cr}.
\end{equation}
The solution of (\ref{53}) is, therefore,
\begin{equation}
\label{64}
n_j(t)\approx n_{j0}\,e^{-\Gamma_j t}\qquad(j=1,2).
\end{equation}
In this case, the levels decouple from each other, and each one of
them decays with its own decay rate. The coupling to the heat bath
is so weak that the system is effectively insulated.

Now we are in position to answer the questions posed in this section:

(A) If the system is overdamped (i.e., $\Gamma\gg\Gamma_1$, $\Gamma_2$),
it decays as indicated in (\ref{50}). The analysis of the
overdamped case also explains why the decay rate is given by (\ref{50}),
instead of (\ref{52}). 
(In some sense, that is why one can observe a phenomenon like the
Stark effect. In the presence of a constant external electric
field, the potential which binds an electron to an atom becomes
unbounded from below, and so the atomic energy levels become
metastable, with very large lifetimes if the external electric
field is small compared to the field of the nucleus. 
The role of the heat bath is played here
by the quantized electromagnetic field; it is the coupling of the
atom to it that causes the excited states of the atom, otherwise
stationary, to decay to the ground state. Even for a highly
excited atom, for which the natural lifetime is relatively high
(some milliseconds), the situation is well described by saying that
the atom is overdamped, since the ionization (decay) rate is very
small if the external electric field is small compared to the
intratomic field.)

(B) On the other hand, if the system is underdamped 
($\Gamma\ll\Gamma_1$, $\Gamma_2$), although the decay is still
described by Eq.\ (\ref{36}), Eq.\ (\ref{49}) is possibly not
valid. The reason is that, under such a
condition, the system does not ``know'' the temperature of the 
heat bath. It would have decayed before it could thermalize.
In practice, as already argued, the excited states would depopulate
much sooner than the ``ground state,'' and so
the decay rate would be very
insensitive to the temperature of the heat bath. 

\section{Thermodynamics of metastable systems: a brief digression}
\label{VI}

Finally, we would like to make a brief digression on a point
where we think there is some confusion in the literature.
It concerns the thermodynamics of metastable systems.
As an example of such a system, let us consider a particle 
interacting with the potential defined in Section \ref{II}.
Its partition function is defined as 
\begin{eqnarray}
\label{65}
{\cal Z}&\equiv&{\rm Tr}\,e^{-\beta H} \nonumber \\
&=&\int_0^{\infty}dk\,e^{-\beta k^2/2}\,
\int_0^{\infty}dx\,\varphi^*(k,x)\,\varphi(k,x),
\end{eqnarray}
or, alternatively, as
\begin{equation}
\label{66}
{\cal Z}=\int_0^{\infty}\rho(E)\,e^{-\beta E}\,dE.
\end{equation}
Clearly, the spectral density $\rho(E)$ is given, in this case, 
by\cite{delta}
\begin{equation}
\label{67}
\rho(E)=\delta(0)\,\frac{dk}{dE}=\frac{\delta(0)}{\sqrt{2E}}.
\end{equation}
We conclude, therefore, that there is no sign of metastability 
in the partition function. Now, let us try to define a ``partition
function inside the well;'' since this is not a fundamental
concept, more than one definition is possible.
One such definition is inspired by (\ref{65}); restricting the
$x$-integration to the interval $[0,a]$, we have
\begin{equation}
\label{68}
Z_1=\int_0^{\infty}dk\,e^{-\beta k^2/2}\,
\int_0^{a}dx\,\varphi^*(k,x)\,\varphi(k,x).
\end{equation}
If we make the same approximations we made in Section \ref{III}, we can
reexpress $Z_1$ as in (\ref{66}), but now with a spectral density
given by
\begin{equation}
\label{69}
\rho_1(E)\approx\sum_n\frac{1}{\pi}\,\frac{\Gamma_n/2}
{(E-\epsilon_n)^2+\Gamma_n^2/4}.
\end{equation}
This kind of spectral density\cite{Patra,Boyan1} 
contains some dynamical information --- resonant levels and decay rates.
If the latter are small enough, the Lorentzians in (\ref{69})
can be approximated by delta functions, and we obtain, therefore,
\begin{equation}
\label{70}
Z_1\approx\sum_n e^{-\beta\epsilon_n}\equiv e^{-\beta F_1}.
\end{equation}
If the coupling with the heat bath is strong (in the sense of 
Section \ref{V}),
$F_1$ can be interpreted \cite{Langer1,Langer2} as the free energy of
the metastable phase. One should not confuse\cite{Boyan1}  
this ``free energy'' with the
true equilibrium free energy, ${\cal F}=-(1/\beta)\,\ln {\cal Z}$.

Another possible definition of the ``partition function inside the
well'' is inspired in the complex eigenenergies of Gamow's method:
\begin{equation}
\label{71}
Z_2\equiv\sum_n e^{-\beta(\epsilon_n-i\Gamma_n/2)}\equiv e^{-\beta F_2}.
\end{equation}
This definition, although almost identical to (\ref{70}), pre\-sents
a new and interesting feature: the free energy $F_2$ of the metastable phase
is complex! Its real part is essentially equal to $F_1$, and so has
thermodynamical content, but its imaginary part provides dynamical
information: if the $\Gamma_n$'s are small, it is easy to show that
\begin{equation}
\label{72}
{\rm Im}\,F_2\approx -\frac{i}{2}\,\langle\Gamma\rangle,
\end{equation}
where $\langle\Gamma\rangle$ is the thermal average of the
$\Gamma_n$'s, as defined in (\ref{50}). That is the reason why
definition (\ref{71}) is so popular in the literature, although
it is no more fundamental than definition (\ref{68}). 

\section{Conclusion}

In Section \ref{III} we showed that decaying states,
although plagued by the exponential catastrophe, give a fairly
good description of the decay of a metastable state, provided
some conditions are satisfied. In fact, the main result
of this paper is that one can compute the decay rate solving
the time independent Schr\"odinger equation subject to the ``outgoing
wave boundary condition,'' Eq.\ (\ref{9}). This is far from
being a trivial result, 
since the corresponding eigenstates
are nonphysical. The ``effectiveness'' of the decaying states
in describing the decay may be understood by noticing\cite{Skib} 
that they are good approximate 
solutions to the time-dependent Schr\"odinger
equation, although nonuniform ones (i.e., they are not valid
in the entire range of values of $t$ and $x$). 

In Section \ref{V} we examined another ``well known'' result, that the
decay rate of a metastable system in contact with a heat bath is
given by Eq.\ (\ref{50}). Although we have used a very simple
toy model to discuss this point, we believe it contains
the essential physics of the phenomenon, at least in the two
limiting cases we studied in some detail. The important lesson
to be learned here is that Eq.\ (\ref{50}) is
correct (at least in first approximation), provided the condition
of ``overdamping'' is satisfied.\cite{Legget,Boyan2}
At low temperatures, where only
the lowest lying decaying states take part in the process, it is
an easy matter to verify if it is so. However,
as the temperature increases, decaying states of higher energy
are excited and begin to play an increasingly important role in
the overall decay. Since the decay rates $\Gamma_n$ 
become larger with $n$, the overdamping condition eventually
fails to be satisfied by states actively involved in the
process of decay. Thus, one should expect deviations from
the decay rate given by Eq.\ (\ref{50}).
Another source of deviations, not taken into account
in our toy model, is a possible renormalization of the complex
energies $E_n$, caused by the interaction of the system
with the heat bath. This may affect Eq.\ (\ref{50})
even at low temperatures, for obvious reasons.\cite{Weiss} 

\acknowledgments

We thank Pavel Exner for pointing out Ref.~5, and Gernot Muenster for 
pointing out Refs.~9 and 10.
This work had financial support from CNPq, FINEP, CAPES and FUJB/UFRJ.



\begin{references}
\bibitem{initial} It can be
shown\cite{CSM,FGR} that initially the decay is slower
than exponential; more precisely, $dP/dt=0$ when $t=0$.
\bibitem{eigenvector} To find the eigenvectors one must, in fact, 
solve (\ref{55}) to first 
order in $\Gamma$: $\lambda_1\approx\Gamma_1+e^{-\beta E}\Gamma$,
$\lambda_2\approx\Gamma_2+\Gamma$.
\end{references}
\end{document}